\documentclass[letterpaper,12pt]{article} 
\usepackage{latexsym}
\usepackage{amsmath,amsfonts}

\providecommand{\e}{\mathrm{e}}
\providecommand{\bvec}[1]{{\mathbf{#1}}}
\providecommand{\sint}[1]{\int_{\partial AdS} d^d{#1}\,}
\providecommand{\kint}{\int\frac{d^dk}{(2\pi)^d}\,}

\providecommand{\tGamma}{\tilde\Gamma}
\providecommand{\tR}{\tilde R}
\providecommand{\tg}{\tilde g}
\providecommand{\hh}{\hat h}

\title{The Graviton in the AdS-CFT correspondence: \\
Solution via the Dirichlet boundary value problem} 
\author{W.~M\"uck\thanks{Email address:
wmueck@sfu.ca}\ ~and K.~S.~Viswanathan\thanks{Email address:
kviswana@sfu.ca}\\ \small Department of Physics, Simon Fraser
University, Burnaby, B.C., V5A 1S6 Canada}

\begin{document} 
\maketitle 
\begin{abstract} 
Using the AdS-CFT correspondence we calculate the two point function of 
CFT energy momentum tensors. The AdS gravitons are considered by
explicitely solving the Dirichlet boundary value problem for $x_0=\epsilon$.
We consider this treatment as complementary to exising work, with which we
make contact.
\end{abstract} \newpage

% for draft (line spacing) 
\renewcommand{\baselinestretch}{1.2}
\normalsize

\section{Introduction} 
\label{intro}
In recent months much has been written about the so-called AdS-CFT 
correspondence. It is a special case of the holographic connection between 
bulk and boundary theories \cite{Susskind} and has attracted much
attention because it presumably links two important theories, namely 
Type IIB supergravity and the large $N$ limit of
${\mathcal{N}}=4$ Super Yang Mills Theory \cite{Maldacena}.  
The principle behind this correspondence is most often stated as
follows \cite{Witten}: 
\begin{equation}
\label{intro:princ}
  Z_{AdS}[\phi_0] = \int_{\phi_0} \mathcal{D}\phi\, \exp(-I[\phi])
  \equiv Z_{CFT}[\phi_0] = \left\langle \exp\left(\sint{x} 
  \mathcal{O} \phi_0]\right) \right\rangle.  
\end{equation}
The path integral over AdS fields on the l.h.s.\ is calculated under the
restriction that the field $\phi$ satisfy a Dirichlet boundary condition 
on the boundary $\partial AdS$, and the corresponding boundary value 
$\phi_0$ acts as a current coupling to the CFT field. The path integal 
is, of course, redundant in the classical approximation. 

However, the prescription \eqref{intro:princ} 
is not applicable as it stands, but must be 
''regularized''. The reason is that fields on anti-de Sitter space 
(except some notable exceptions) do not extend to the boundary, which 
means that the Dirichlet boundary value problem is not well defined. 
For regularization, one modifies the
theory such that for any finite value of some parameter
$\epsilon$ the prescription be well defined and one takes the limit 
$\epsilon\to0$ at the end. One possible scheme, which has been shown to
be consistent, is to shift the boundary to the interior of the AdS
bulk and shift it back after the calculations. For most calculations, 
notably those of CFT three and higher point functions, which depend on 
the bulk rather than only on the boundary behaviour of the AdS fields,
the limit can be taken as an intermediate step, which greatly simplifies the
calculations. However, as was pointed out in \cite{Freedman}, 
one must strictly follow the regularization
procedure for CFT two point functions in order for the results to be
consistent (Ward identities). A different method of regularization 
can be found in \cite{Volovich2,Solodukhin}.

The regularization scheme described above has been consequently applied to the
treatment of scalar \cite{Mueck1}, vector and spinor \cite{Mueck2},
Rarita-Schwinger fields (gravitinos) \cite{Corley}, the transverse
traceless part of gravitons \cite{Arutyunov} and antisymmetric tensor
fields \cite{Arutyunov2}.
Gravitons are one of the ''notable exceptions,'' 
where the limit can indeed be taken as an intermediate step also 
when calculating the CFT two point function. They have been 
considered in \cite{Liu} using this fact. Despite our results being identical 
with those already known, we feel that a full treatment of 
gravitons within this regularization procedure should complement the other 
fields. Thus, we shall in this
paper present our calculation of the two point function of CFT energy 
momentum tensors, which couple to the AdS gravitons. We will work in the
radiation gauge $h_{\mu0}=0$ and use the time slicing formalism, which
particularly suits the problem. Then, we shall make contact with the
existing work \cite{Liu} by explicitely calculating the necessary gauge
transformation.

The AdS-CFT correspondence is a field, where progress is being made quickly, 
as the sheer number of recent articles shows. Let us therefore mention
only a few. Two and three point functions were calculated in
\cite{Gubser,Witten,Henningson,Mueck1,Freedman,Liu,Chalmers1,Mueck2,Solodukhin,Lee,Hoker1,Corley}.
Contributions to four point functions can be found in
\cite{Mueck1,Liu2,Freedman2,Brodie,Hoker2,Chalmers2}. 

\section{Notations and Background}
\label{not}
The representation of Euclidean anti-de Sitter space most often used in
the AdS-CFT context is the ''upper half space,'' $x^i\in\mathbb{R}$ 
($i=1,\ldots,d$), $x_0>0$, with the metric
\begin{equation}
\label{not:ads}
  ds^2 = x_0^{-2} dx^\mu dx^\mu,
\end{equation}
where we sum over the indices $\mu=0,1,\ldots,d$. The natural boundary of
this space is given by $x_0=0$, but in view of what we said above 
we shall consider the hypersurface $x_0=\epsilon$ as the boundary 
and take the limit $\epsilon\to0$ at the end. We conveniently split up the
coordinates of AdS$_{d+1}$ as $x=(x_0,\bvec{x})$, where $\bvec{x}$ has
components $x^i$. 

Given this notation, the idea of
using the initial value formulation of gravity with the time slices defined
by $x_0=const.$ hypersurfaces is very intriguing. 
 
Thus, let us begin with a review of the basic geometric relations
for immersed hypersurfaces \cite{Eisenhart}. Let a hypersurface be 
defined by the functions $X^\mu(x^i)$ and let
$\tg_{\mu\nu}$ and $g_{ij}$ be the metric tensors of the imbedding
manifold and the hypersurface, respectively. Raising and lowering of
indices is, for the time being, performed covariantly using these
metrics. The tangents $\partial_i X^\mu$ and normal $N^\mu$ of the
hypersurface satisfy the following orthogonality relations:
\begin{align}
\label{not:ort1}
  \tg_{\mu\nu} \partial_i X^\mu \partial_j X^\nu &= g_{ij}, \\
\label{not:ort2}
  \partial_i X^\mu N_\mu &=0,\\
\label{not:ort3}
  N_\mu N^\mu &=1.
\end{align}
We shall in the sequel use a tilde to label quantities relating to the
imbedding $d+1$ dimensional manifold and leave those relating to the
hypersurface unadorned. Moreover, we use the symbol $D$ to denote a
covariant derivative with respect to whatever indices follow.
Then, there are the equations of Gauss and
Weingarten, which define the second fundamental form $H_{ij}$ of the
hypersurface,
\begin{align}
\label{not:gauss1}
  D_i \partial_j X^\mu &\equiv \partial_i\partial_j X^\mu -\Gamma^k_{ij}
  \partial_k X^\mu +\tGamma^\mu_{\lambda\nu}\partial_i X^\lambda
  \partial_j X^\nu = H_{ij} N^\mu,\\
\label{not:gauss2}
  D_i N^\mu &\equiv \partial_i N^\mu + \tGamma^\mu_{\lambda\nu} \partial_i
  X^\lambda N^\nu = - H_i^j \partial_j X^\mu.
\end{align}
The second fundamental form describes the extrinsic curvature of the 
hypersurface and is related to the intrinsic curvature by another
equation of Gauss,
\begin{equation}
\label{not:gauss3}
  \tR_{\mu\nu\lambda\rho} \partial_i X^\mu \partial_j X^\nu \partial_k
  X^\lambda \partial_l X^\rho = R_{ijkl} + H_{il}H_{jk} - H_{ik}
  H_{jl}.
\end{equation}
Using the Eqns.\ \eqref{not:gauss1}-\eqref{not:gauss3} one can show
that the curvature $\tR$ can be written as 
\begin{equation}
\label{not:tR}
  \tR = R - H_{ij} H^{ij} + H^2 + 2 D_\mu \left(N^\nu D_\nu N^\mu -
  N^\mu D_\nu N^\nu \right),
\end{equation} 
where $H=H^i_i$ is the trace of the second fundamental form.

Let us consider the action \cite{Liu,Arutyunov}
\begin{equation}
\label{grav:action1}
  I = - \int d^{d+1}x \sqrt{\tg} \left[ \tR +d(d-1) \right] + \int d^dx
  \sqrt{g} \left( 2H +2(d-1) \right].
\end{equation}
It consists of the Einstein-Hilbert action with a cosmological constant, 
the Gibbons-Hawking term and a cosmological boundary term, which was
justified in \cite{Liu,Arutyunov}.
Using Eqn.\ \eqref{not:tR} one realizes that the total derivative
cancels the Gibbons-Hawking term. Moreover, converting the cosmological
boundary term into a bulk integral by inserting $N_\mu N^\mu=1$ one
arrives at the expression
\begin{equation}
\label{grav:action2}
  I = - \int d^{d+1}x \sqrt{\tg} \left[ R-H_i^j H^i_j +H^2 +d(d-1) +
  2(d-1) H \right].
\end{equation}
The advantages of this representation of the action become pronounced 
when using the time slicing formalism 
and working in radiation gauge $h_{\mu0}=0$.

In the time slicing formalism \cite{Wald,MTW} we consider the bundle 
of immersed hypersurfaces 
defined by $X^0=const.$, whose tangent vectors are given by 
$\partial_i X^0=0$ and 
$\partial_i X^\mu =\delta_i^\mu$ ($\mu=1,2,\ldots d$). One conventionally 
splits up the metric as 
\begin{align}
\label{not:tg}
  \tg_{\mu\nu} &= \begin{pmatrix} 
      n^i n_i +n^2 & n_k\\
      n_i        &g_{ik} \end{pmatrix},\\
\intertext{whose inverse is given by}
\label{not:tginv}
  \tg^{\mu\nu} &= \frac{1}{n^2} \begin{pmatrix}
      1    &   -n^i \\
      -n^k & n^2 g^{ki} +n^k n^i \end{pmatrix}
\end{align}
and whose determinant is $\tg = n^2 g$. The quantities $n$ and $n^i$ are 
called the lapse function and shift vector, respectively \cite{MTW}. 
The normal vectors $N^\mu$ satisfying \eqref{not:ort2} and \eqref{not:ort3}
are then given by  
\begin{equation}
\label{not:N}
  N_\mu = (-n,\bvec{0}),\quad N^\mu = \frac{1}{n}(-1, n^i),
\end{equation}
where the sign has been chosen such that the normals point outwards on
the boundary ($n>0$ without loss of generality). 
Then, using Eqn.\ \eqref{not:gauss2} one obtains
\begin{equation}
\label{not:H}
  H_{ij}= \frac{1}{2n} ( -D_i n_j - D_j n_i + g_{ij,0} ).
\end{equation}

The equations of motion for $n$ and $n^i$ obtained from the action
\eqref{grav:action2} are given by
\begin{align}
\label{not:con1}
  R+d(d-1)-H^2 + H_i^j H^i_j &=0\\
\intertext{and}
\label{not:con2}
  D_i H - D_j H^j_i &=0,
\end{align}
respectively. However, these two equations represent only constraint 
equations in the initial value formulation of gravity. 

The time slicing formalism considered so far is very useful for our 
purposes when using
the radiation gauge $h_{\mu0}=0$, which directly translates into
\begin{equation}
\label{not:radgauge}
  n=\frac{1}{x_0}, \quad n^i=0.
\end{equation}
Then, the remaining physical degrees of freedom are the metric
components $g_{ij}$. 
Writing as usual
\begin{equation}
\label{not:g}
  g_{ij} = \bar g_{ij} + h_{ij},
\end{equation}
where $\bar g_{ij} = x_0^{-2} \delta_{ij}$ is the background metric, we
shall use $h^i_j = \bar g^{ik} h_{kj}$ as the graviton fields and from
now on raise and lower the latin indices with the Euclidean metric.

\section{Free Gravitons and Two Point Function}
\label{grav}
Linearizing the constraint equations \eqref{not:con1} and
\eqref{not:con2} one obtains the relations
\begin{align}
\label{grav:lcon1}
   \Box h -\partial_i \partial^j h^i_j + \frac{1-d}{x_0} \partial_0 h
  &=0\\
\intertext{and}
\label{grav:lcon2}
  \partial_0 \left(\partial_i h -\partial_j h^j_i \right) &=0,
\end{align}
respectively, where $\Box=\partial_i\partial^i$. 
Substituting the gauge conditions \eqref{not:radgauge} into the action
\eqref{grav:action2} and expanding to second order in $h^i_j$ one finds
\begin{equation}
\label{grav:action3}
  I = \frac{1}{4} \int d^{d+1}x\, x_0^{1-d} \left( h^k_{j,i} h^{j,i}_k 
  - h_{,i} h^{,i} - 2 h^k_{j,i} h^{i,j}_k + 2 h_{,i} h^{i,j}_j 
  + h^i_{j,0} h^j_{i,0} - h_{,0} h_{,0}  \right). 
\end{equation}
Zeroth and first order terms do not appear. The equation of motion for
$h_i^j$ derived from Eqn.\ \eqref{grav:action3}, suitably combined with
the constraints \eqref{grav:lcon1} and \eqref{grav:lcon2} yields the
equation
\begin{equation}
\label{grav:hmot}
  \partial_0^2 h^i_j + \Box h^i_j + \frac{1-d}{x_0} \partial_0 h^i_j
  -\frac{1}{x_0} \partial_0h \delta^i_j +\partial^i\partial_j h
  -\partial^i\partial_l h^l_j -\partial_j\partial^l h^i_l =0.
\end{equation}
The Eqns.\ \eqref{grav:lcon1}, \eqref{grav:lcon2} and \eqref{grav:hmot}
are identical with the ones found by Arutyunov and Frolov \cite{Arutyunov}. 
We shall now solve the
Dirichlet boundary value problem of these Eqns.\ and use the result to
calculate the two point function of CFT energy momentum tensors.

Comparing the trace of Eqn.\ \eqref{grav:hmot} with the constraint
\eqref{grav:lcon1} yields the simple equation for $h$,
\begin{equation}
\label{grav:heq}
  \partial_0^2 h -\frac{1}{x_0} \partial_0 h =0,
\end{equation}
whose solution is given by
\begin{equation}
\label{grav:h1}
  h = \frac12 (x_0^2 -\epsilon^2) b(\bvec{x}) + a(\bvec{x}).
\end{equation}
Solving the constraint \eqref{grav:lcon2} one then obtains
\begin{equation}
\label{grav:hdiv1}
  \partial_j h^j_i = \frac12 (x_0^2 -\epsilon^2) \partial_i b(\bvec{x}) 
  + a_i(\bvec{x}).
\end{equation}
Substituting Eqns.\ \eqref{grav:h1} and \eqref{grav:hdiv1} into
Eqn.\ \eqref{grav:lcon1} one finds 
\begin{equation}
\label{grav:b} 
  b = \frac{1}{d-1} (\Box a - \partial^i a_i).
\end{equation}
Thus, writing 
\begin{align}
\label{grav:h2}
  h &= \frac{x_0^2 -\epsilon^2}{2(d-1)} (\Box a - \partial^i a_i) +a,\\
\label{grav:hdiv2}
  \partial_j h^j_i &= \frac{x_0^2 -\epsilon^2}{2(d-1)} \partial_i
  (\Box a - \partial^j a_j) +a_i,
\end{align}
we have found the trace and divergence of $h^i_j$ as functions of their 
boundary values $a=h|_{x_0=\epsilon}$ and 
$a_i=\partial_j h^j_i|_{x_0=\epsilon}$. 

We now substitute the solutions \eqref{grav:h1} and \eqref{grav:hdiv1} into
the equation of motion \eqref{grav:hmot} and obtain the inhomogeneous
differential equation
\begin{equation}
\label{grav:hmot2}
  \partial_0^2 h^i_j + \Box h^i_j + \frac{1-d}{x_0} \partial_0 h^i_j
  = \delta^i_j b - \partial^i\partial_j a + \partial_j a^i +\partial^i
  a_j +\frac12 (x_0^2-\epsilon^2) \partial^i\partial_j b.
\end{equation}
The homogeneous part of Eqn.\ \eqref{grav:hmot2} is identical with the 
equation of motion of a massless scalar field \cite{Mueck1}, whose
solution we can take over. For the particular solution we make the
ansatz
\begin{equation}
\label{grav:hp}
  h^{(p)i}_j(x) = \frac12 (x_0^2-\epsilon^2) c^i_j(\bvec{x}) +
  d^i_j(\bvec{x})
\end{equation}
and obtain 
\begin{align}
\label{grav:c}
  c^i_j &= \frac{\partial^i\partial_j}\Box  b\\
\intertext{and}
\label{grav:d}
  d^i_j &= \delta^i_j \frac1\Box b -\frac{\partial^i\partial_j}\Box a
  +\frac{\partial_j}\Box a^i + \frac{\partial^i}\Box a_j + 
  (d-2) \frac{\partial^i \partial_j}{\Box^2} b.
\end{align}
Taking the trace and divergence of the particular solution
\eqref{grav:hp} and comparing with Eqns.\ \eqref{grav:h1} and \eqref{grav:hdiv1}
one finds $h^{(p)} = h$ and $\partial_j h^{(p)j}_i =
\partial_j h^j_i$, respectively, which means that the homogeneous solution 
of Eqn.\ \eqref{grav:hmot2} must be traceless and transversal. 
Combining the homogeneous and inhomogeneous solutions and using the
relations \eqref{grav:c}, \eqref{grav:d} and \eqref{grav:b} we finally obtain
the free graviton field in radiation gauge,
\begin{multline}
\label{grav:hsol}
  h^i_j(x) = \kint \e^{-i\bvec{k}\cdot\bvec{x}} \left[
  \bar{\hh^i_j} \left(\frac{x_0}{\epsilon}\right)^\frac{d}2
  \frac{K_{\frac{d}2}(kx_0)}{K_{\frac{d}2}(k\epsilon)}
  -\frac{x_0^2-\epsilon^2}{2(d-1)} \left(k^ik_j\hh-\frac{k^ik_jk^kk_l}{k^2}
  \hh^l_k \right) \right.\\
  \left. +\frac{k_jk^l}{k^2}\hh^i_l +\frac{k^ik_l}{k^2}\hh^l_j
  -\frac{k^ik_jk^kk_l}{k^4} \hh^l_k
  +\frac1{d-1}\left(\delta^i_j-\frac{k^ik_j}{k^2}\right)
  \left(\hh-\frac{k^kk_l}{k^2}\hh^l_k\right)\right],
\end{multline}
where $\hh^i_j(\bvec{k})$ is the Fourier transform of the boundary data
$h^i_j|_{x_0=\epsilon}$. Moreover, setting $x_0=\epsilon$ in Eqn.\
\eqref{grav:hsol} one finds
\begin{equation}
\label{grav:hbar}
  \bar{\hh^i_j} = \hh^i_j -\frac{k_jk^l}{k^2}\hh^i_l -\frac{k^ik_l}{k^2}\hh^l_j
  +\frac{k^ik_jk^kk_l}{k^4} \hh^l_k
  -\frac1{d-1}\left(\delta^i_j-\frac{k^ik_j}{k^2}\right)
  \left(\hh-\frac{k^kk_l}{k^2}\hh^l_k\right),
\end{equation}
which is the traceless transversal part of $h^i_j$ in momentum space
\cite{Arutyunov}. 

Let us continue by calculating the two-point function of the CFT
energy momentum tensor $T_i^j(\bvec{x})$, which couples to $h^i_j$ on the
boundary. We shall be interested in its 
non-local parts only. First, using the equations of
motion \eqref{grav:hmot} the action \eqref{grav:action3} takes the value
\begin{equation}
\label{grav:action4}
  I=\frac14 \int d^dx\, \epsilon^{1-d} \left( -h^j_i \partial_0 h^i_j +
  h\partial_0 h \right),
\end{equation}
which tells us that the normal derivative of the field on the boundary,
$\partial_0 h^i_j|_{x_0=\epsilon}$, contains all the information needed.
From Eqn.\ \eqref{grav:hsol} one finds
\begin{equation}
\label{grav:dh1}
  \partial_0 h^i_j|_\epsilon = \kint \e^{-i\bvec{k}\cdot\bvec{x}} \left[
  \bar{\hh^i_j} \frac{k}\epsilon \frac{d}{dk}
  \ln\left((k\epsilon)^\frac{d}2 K_\frac{d}2 (k\epsilon) \right) 
  - \frac{\epsilon}{d-1} \left(k^ik_j\hh
  -\frac{k^ik_jk^kk_l}{k^2}\hh^l_k \right)\right].
\end{equation}
Taking the trace of Eqn.\ \eqref{grav:dh1} one realizes that $\partial_0
h|_\epsilon$ is a local expression and thus the corresponding term in Eqn.\ 
\eqref{grav:action4} does not contribute to the non-local part of the 
two-point function. We now expand the modified Bessel function using the
formula \cite{Gradshteyn}
\begin{equation}
\label{grav:bessel}
  z^\alpha K_\alpha = 2^{\alpha-1} \Gamma(\alpha)
  \left[1+\frac{z^2}{4(1-\alpha)}
  -\frac{\Gamma(1-\alpha)}{\Gamma(1+\alpha)}
  \left(\frac{z}2\right)^{2\alpha} +\cdots \right],
\end{equation}
where the ellipsis indicates either terms of order $z^4$ and higher,
which yield local terms in the integral \eqref{grav:dh1} as well as terms of
order $z^{2\alpha+2}$ or higher, which are negligible in the
$\epsilon\to0$ limit. In contrast to the scalar and vector fields 
\cite{Mueck1,Mueck2} we have to keep the $z^2$ term, which gives a nonlocal 
contribution together with the $k^{-4}$ term contained in
$\bar{\hh^i_j}$. However, one finds that this term cancels the nonlocal 
term proportional to $\epsilon$ in Eqn.\ \eqref{grav:dh1}. Hence, neglecting 
all local terms we find
\begin{equation}
\label{grav:dh2}
\begin{split}
  \partial_0h^i_j|_\epsilon =& \int d^dy\, h^r_s(\bvec{y}) \kint
  \e^{-i\bvec{k}\cdot(\bvec{x}-\bvec{y})} \left(-\epsilon^{d-1} 2^{-d}
  d\frac{\Gamma\left(1-\frac{d}2\right)}{\Gamma\left(1+\frac{d}2\right)} 
  k^d \right)\\
  &\times\left[\frac12\delta^i_r\delta^s_j +\frac12\delta^{is}\delta_{jr}
  -\frac{k_jk^s}{2k^2}\delta^i_r -\frac{k_jk_r}{2k^2}\delta^{is}
  -\frac{k^ik_r}{2k^2}\delta^s_j
  -\frac{k^ik^s}{2k^2}\delta_{jr}\right. \\
  &\left.+\frac{d-2}{d-1} \frac{k^ik_jk^sk_r}{k^4}
  +\frac1{d-1}\left(-\delta^i_j\delta^s_r +\frac{k_rk^s}{k^2}\delta^i_j
  +\frac{k^ik_j}{k^2}\delta^s_r \right)\right].
\end{split}
\end{equation}
Finally, carrying out the $k$ integration and substituting the result
into Eqn.\ \eqref{grav:action4} yields
\begin{equation}
\label{grav:action5}
  I= -\frac{\kappa d}4 \int d^dx d^dy\, 
  \frac{h^j_i(\bvec{x}) h^r_s(\bvec{y})}{|\bvec{x}-\bvec{y}|^{2d}} 
  \left[\frac12\left(J^i_r(\bvec{x}-\bvec{y})J^s_j(\bvec{x}-\bvec{y})
  +J^{is}(\bvec{x}-\bvec{y})J_{jr}(\bvec{x}-\bvec{y})\right)
  -\frac1d \delta^i_j\delta^s_r\right],
\end{equation}
where we introduced 
\begin{equation}
\label{grav:J}
  J^i_j(x) = \delta^i_j -2\frac{x^i x_j}{x^2}
\end{equation}
and 
\begin{equation}
\label{grav:kappadef}
  \kappa= \frac{d+1}{d-1} 
  \frac{\Gamma(d)}{\pi^\frac{d}2 \Gamma\left(\frac{d}2\right)}.
\end{equation}
From Eqn.\ \eqref{grav:action5} we can read off the two-point function 
\begin{equation}
\label{grav:2point}
  \left\langle T^i_j(\bvec{x})T^s_r(\bvec{y}) \right\rangle
  = \frac{\kappa d}{2|\bvec{x}-\bvec{y}|^{2d}} 
  \left[\frac12\left(J^i_r(\bvec{x}-\bvec{y})J^s_j(\bvec{x}-\bvec{y})
  +J^{is}(\bvec{x}-\bvec{y})J_{jr}(\bvec{x}-\bvec{y})\right)
  -\frac1d \delta^i_j\delta^s_r\right],
\end{equation}
which is just as expected and coincides with the one found by Liu and
Tseytlin \cite{Liu}. 

\section{Transformation to de Donder Gauge}
\label{gauge}
Having found the free graviton field in radiation gauge, it is in
principle straightforward to obtain solutions in any other gauge,
the covariant gauges being particularly important.
Performing a gauge transformation
\begin{equation}
\label{gauge:trafo}
  {h'}^\mu_\nu = h^\mu_\nu +\nabla^\mu \xi_\nu +\nabla_\nu \xi^\mu,
\end{equation}
where $\nabla$ is the covariant derivative with respect to the
background metric, we want to impose the de~Donder gauge condition
\begin{equation}
\label{gauge:cond}
  \nabla_\mu \left({h'}^\mu_\nu -\frac12 \delta^\mu_\nu h'\right) =0,
\end{equation}
which is the standard gauge when dealing with linearized gravity.
Eqns.\ \eqref{gauge:trafo} and \eqref{gauge:cond} imply that we have to
solve the inhomogeneous differential equation
\begin{equation}
\label{gauge:xieqn}
  \nabla_\mu \nabla^\mu \xi_\nu + \bar R_\nu^\mu \xi_\mu 
  = -\nabla_\mu h^\mu_\nu + \frac12 \partial_\nu h,
\end{equation}
with the boundary condition $\xi_\mu|_{x_0=\epsilon}=0$ in order to preserve the
boundary data of the graviton field. In Eqn.\ \eqref{gauge:xieqn}
$\bar R^\mu_\nu$ is the background Ricci tensor. Keeping in mind 
that $h^\mu_\nu$ is in radiation gauge, Eqn.\ \eqref{gauge:xieqn} reads
for the AdS background
\begin{align}
\label{gauge:xi0eq}
  x_0^2(\partial_0^2 +\Box) \xi_0 +(3-d) x_0 \partial_0 \xi_0
  -(3d-1)\xi_0 + 2x_0 \partial^i \xi_i &= \frac12 \partial_0 h
  -\frac1{x_0}h,\\
\label{gauge:xiieq}
  x_0^2(\partial_0^2 +\Box) \xi_i +(3-d) x_0 \partial_0 \xi_i
  -2d\xi_i - 2x_0 \partial_i \xi_0 &= \frac12 \partial_i h - \partial_j
  h^j_i.
\end{align}
Going into momentum space and using the solution \eqref{grav:hsol} 
we find the particular solution of Eqns.\
\eqref{gauge:xi0eq} and \eqref{gauge:xiieq} as 
\begin{equation}
\label{gauge:xi0p}
  \xi^{(p)}_0 (\bvec{k},x_0)
  = \frac1{2(d-1)x_0} \left(\hh -\frac{k^ik_j}{k^2} \hh^j_i
  \right)
\end{equation}
and
\begin{equation}
\label{gauge:xiip}
\begin{aligned}
  \xi^{(p)}_i (\bvec{k},x_0) 
  =& -\frac{i}{k^2x_0^2} \left(k_j \hh^j_i -\frac{k_i
  k^jk_l}{k^2} \hh^l_j \right)
  +\frac{ik_i}{2(d-1)k^2x_0^2} \left(\hh- d\frac{k^jk_l}{k^2}\hh^l_j
  \right)\\
  &+\frac{ik_i}{4(d-1)} \left(\hh -\frac{k^jk_l}{k^2} \hh^l_j \right)
  \left(1-\frac{\epsilon^2}{x_0^2} \right).
\end{aligned}
\end{equation}
The terms in Eqn.\ \eqref{gauge:xiip} have been arranged such that the
first and second terms are transversal and longitudinal, respectively, 
and the third term vanishes on the boundary. 
It is interesting to note that, after a gauge transformation using  
$\xi^{(p)}_\mu$, only the traceless transversal part of $h^i_j$
survives, while the components $h^0_\mu$ remain zero. Thus, we found a
graviton solution satisfying the traceless transversal gauge
\cite{Wald}, but this solution does not solve the Dirichlet boundary
value problem, because also its boundary values are constrained to be
traceless and transversal.

Turning to the homogeneous part of the Eqns.\ \eqref{gauge:xi0eq} and
\eqref{gauge:xiieq} we realize that there exists a solution with
$\xi_0=0$ and $k^i \xi_i=0$. Hence, the coefficient of this solution 
must be chosen such that it cancels the first term on the r.h.s.\ of Eqn.
\eqref{gauge:xiip} on the boundary. On the other hand, the homogeneous 
solutions containing the longitudinal part of $\xi_i$ 
must cancel $\xi^{(p)}_0$ and the
second term on the r.h.s.\ of Eqn.\ \eqref{gauge:xiip} on the boundary.
Skipping the details we find 
\begin{align}
\label{gauge:xi0h}
  \xi^{(h)}_0 (\bvec{k},x_0) = &A x_0^\frac{d}2 K_{\frac{d}2+1}(kx_0)
  + B \left[ \frac1k x_0^\frac{d}2 K_{\beta+1}(kx_0) - \frac{\frac{d}2
  +\beta}{k^2} x_0^{\frac{d}2-1} K_\beta(kx_0) \right]\\
\intertext{and}
\notag
  \xi^{(h)}_i(\bvec{k},x_0) = &\frac{i}{x_0^2k^2} \left(k_jh^j_i
  -\frac{k_ik^jk_l}{k^2}h^l_j \right)
  \left(\frac{x_0}\epsilon\right)^{\frac{d}2+1}
  \frac{K_{\frac{d}2+1}(kx_0)}{K_{\frac{d}2+1}(k\epsilon)}\\
\label{gauge:xiih}
  &+ A \frac{ik_i}{k^2} \left[ kx_0^\frac{d}2 K_\frac{d}2(kx_0) 
  +d x_0^{\frac{d}2-1} K_{\frac{d}2+1}(kx_0) \right] +B \frac{ik_i}{k^2}
  x_0^\frac{d}2 K_\beta(kx_0),
\end{align}
where $\beta=\sqrt{d^2/4 +2d}$. The coefficients $A(\bvec{k})$ and
$B(\bvec{k})$ are uniquely determined by requiring that $\xi^{(p)}_0$
and the second term on the r.h.s.\ of Eqn.\ \eqref{gauge:xiip} be
cancelled on the boundary. 

Let us look at the bulk behaviour of the transformed field
${h'}^\mu_\nu$. In order to obtain it, we need first to determine $A$
and $B$ in the limit $\epsilon\to0$. Using the requirement stated above
and the leading order behaviour of the modified Bessel functions
$K(k\epsilon)$ one obtains 
\begin{align}
\label{gauge:A}
  A|_{\epsilon\to0} &= -\frac{(k/2)^{\frac{d}2+1}}{(d-1)
  \Gamma(d/2+1)} \left(\frac{\hh}d
  -\frac{k_ik^j}{k^2}\hh^i_j \right),\\
\label{gauge:B}
  B|_{\epsilon\to0} &= -\frac{\epsilon^{\beta-\frac{d}2} k^{\beta+2}\hh}{
  2^\beta d (\beta-d/2) \Gamma(\beta)}.
\end{align}
However, since $\beta>d/2$, we can set $B=0$ in the transformed field. 
Hence, substituting Eqns.\ \eqref{gauge:xi0p}, \eqref{gauge:xiip},
\eqref{gauge:xi0h}, \eqref{gauge:xiih}, \eqref{gauge:A} and $B=0$ 
into Eqn.\ \eqref{gauge:trafo} one finds after integration
\begin{equation}
\label{gauge:hprime}
  {{h'}^{bulk}}^\mu_\nu(x) =\int d^dy\, 
  \frac{\kappa x_0^d}{(x_0^2+|\bvec{x}-\bvec{y}|^2)^d} J^\mu_k(x-\bvec{y}) 
  J^l_\nu(x-\bvec{y}) P^{kj}_{li} h^i_j(\bvec{y}),
\end{equation}
with the traceless projector
\begin{equation}
\label{gauge:P}
  P^{kj}_{li} = \frac12\left(\delta^k_i\delta^j_l +\delta^{kj}\delta_{li} 
  \right) - \frac1d \delta^k_l \delta^j_i.
\end{equation}
Obviously, Eqn.\ \eqref{gauge:hprime} represents the result found by Liu and
Tseytlin \cite{Liu}. 

\section{Conclusions}
\label{conc}
We have considered the free graviton on AdS background and explicitely
solved the Dirichlet boundary value problem. We took care to carefully
follow the regularization procedure and thus took the limit
$\epsilon\to0$ at the end of the calculation of the CFT two point
function. Our results coincide with earlier results by Liu and Tseytlin
\cite{Liu} as expected. 
The $h_{\mu0}=0$ gauge seems to be useful also for
interacting fields \cite{Chalmers2}. Thus, we hope that our lengthy
explanation of the geometric basics will contribute to the 
understanding of these cases.

\section*{Acknowledgements}
This work was supported in part by NSERC. W.~M.\ gratefully acknowledges
the financial support from Simon Fraser University.

\renewcommand{\baselinestretch}{1}

\end{document}